\documentclass[a4size]{aa}
\usepackage{graphicx,natbib,hyperref}
\usepackage{txfonts}

\newcommand{\Hm}{\langle B\rangle}
\newcommand{\Hz}{\langle B_z\rangle}

\newcommand{\vsi}{v\,\sin i}
\newcommand{\Prot}{P_{\rm rot}}

\newcommand{\Zeeman}{\Delta\lambda_{\rm Z}}

\newcommand{\Feline}{Fe~{\sc ii}~$\lambda\,6149.2$}

\bibpunct[ ]{(}{)}{;}{a}{}{,}

\begin{document}

\title{HD~18078: A very slowly rotating Ap star\\ with an unusual
  magnetic field structure\thanks{Based in part on observations made at
Observatoire de Haute Provence (CNRS), France; at Kitt Peak
National Observatory, National Optical Astronomy Observatory (NOAO
Prop. ID: KP2442; PI: T.~Lanz), which is operated by the Association of Universities
for Research in Astronomy (AURA) under cooperative agreement with the
National Science Foundation; at the
Canada-France-Hawaii Telescope (CFHT) which is operated from the
summit of Maunakea by the National Research Council of Canada, the
Institut National des Sciences de l'Univers of the Centre National de
la Recherche Scientifique of France, and the University of Hawaii; and
at the 6-m telescope BTA of the Special Astrophysical Observatory of
the Russian Academy of Sciences.}}

\author{G.~Mathys\inst{1} 
\and I.~I.~Romanyuk\inst{2} 
\and D.~O.~Kudryavtsev\inst{2}
\and J.~D.~Landstreet\inst{3,4}
\and D.~M.~Pyper\inst{5}
\and S.~J.~Adelman\inst{6}}

\institute{Joint ALMA Observatory \& European Southern Observatory,
  Alonso de Cordova 3107, Santiago, Chile\\\email{gmathys@eso.org}
\and
Special Astrophysical Observatory, Russian Academy of Sciences,
Nizhnii Arkhyz, 369167 Russia 
\and
Armagh Observatory, College Hill, Armagh, BT61 9DG, Northern Ireland,
UK 
\and
Department of Physics \& Astronomy, University of Western Ontario,
London, Ontario N6A 3K7, Canada 
\and
Physics/Astronomy Department, University of Nevada, Las Vegas, 4505
S. Maryland Parkway, Las  Vegas, NV 89154-4002, USA
\and
Department of Physics, The Citadel, 171 Moultrie Street, Charleston,
SC 29409, USA} 

\abstract
{The existence of a significant population of Ap stars with very long
  rotation periods (up to several hundred years) has progressively
  emerged over the past two decades. However, only lower limits of the
  periods are known for most of them because their variations have
  not yet been observed  over a sufficient timebase.}
{We determine the rotation period of the slowly rotating Ap star
  HD~18078 and we derive constraints on the geometrical structure of
  its magnetic field.}
{We combine measurements of the mean magnetic field modulus obtained
  from 1990 to 1997 with determinations of the mean longitudinal
  magnetic field spanning the 1999--2007 time interval to derive an
  unambiguous value of the rotation period. We show that this value is
  consistent with photometric variations recorded in the Str\"omgren
  $uvby$ photometric system between 1995 and 2004. We fit the
  variations of the two above-mentioned field moments with a
  simple model to constrain the magnetic structure.}
{The rotation period of HD~18078 is $(1358\pm12)$~d. The geometrical
  structure of its magnetic field is consistent to first order with
  a colinear multipole model 
  whose axis is offset from the centre of the star.}
{HD~18078 is only the fifth Ap star with a rotation period longer
  than 1000 days for which the exact value of that period (as opposed
  to a lower limit) could be determined. The strong anharmonicity of
  the variations of its mean longitudinal magnetic field and the shift
  between their extrema and those of the mean magnetic field modulus
  are exceptional and indicative of a very unusual magnetic
  structure.}


\maketitle

\titlerunning{HD~18078: a very slowly rotating and unusual Ap star}
\authorrunning{G. Mathys et al.}

\section{Introduction}
\label{sec:intro}
A by-product of systematic investigation of the magnetic fields of Ap
stars whose spectral lines are observationally resolved into their Zeeman
components \citep[and references therein]{1997A&AS..123..353M} is the
finding that the rate of occurrence of long rotation periods in Ap
stars is considerably higher than was previously thought. There are
at present 35 such stars known to have rotation periods $\Prot$ longer
than a month (30 days). There is no doubt  that these slow rotators
represent a significant fraction, of the order of several percent, of
the total population of Ap stars. There are also strong indications
that the longest periods must reach of the order of 300 years, and
even much higher values -- 1000 years  -- seem plausible
\citep{2015ASPC..494....3M}. 

At the other end of the distribution, the fastest rotating Ap stars
have periods of about 0.5~day
\citep[e.g.][]{2002BaltA..11..475A,2004IAUS..215..270M}. The 5 to 6
orders of magnitude spanned by their rotation periods make 
the Ap stars unique on the main sequence. The evolutionary changes of
the rotation periods of Ap stars during their main-sequence lifetimes
are small,  a factor 2 at most
\citep{2006A&A...450..763K,2007AN....328..475H}. Thus period
differentiation must have been completed before Ap stars reach the
main sequence. How this can be achieved in the early stages of
the evolution of stars within a quite limited range of masses ($\sim$1.8 to $\sim$3.0~$M_\odot$) represents a major theoretical
challenge. The mystery of the very long rotation periods is not
primarily one of angular momentum since almost all angular momentum
is already gone at modest periods (10 days). A magnetic stellar
wind can in principle spin a star down indefinitely, but at long
periods it becomes very inefficient. 

\begin{table}
\caption{Mean magnetic field modulus measurements.}
\label{tab:modulus}
\centering
\begin{tabular}{crcl}
\hline\hline\\[-4pt]
HJD&\multicolumn{1}{c}{$\Hm$ (G)}&Observatory&Remarks\\[4pt]
\hline\\[-4pt]
2448166.570&3318&OHP&\\
2448636.341&4415&OHP&\\
2448930.638&3633&OHP&\\
2449286.660&$\la2600$&OHP&unresolved\\ 
2449405.373&$\la2600$&OHP&unresolved\\ 
2449729.883&4014&CFHT&\\
2450055.010&4203&CFHT&\\
2450295.588&3278&OHP&\\
2450346.897&2884&KPNO&\\
2450349.936&3146&KPNO&\\
2450797.471&2682&OHP&\\[4pt]
\hline
\end{tabular}
\end{table}

In the absence of clear theoretical ideas, progress continues to
depend on observational clues; one   limitation 
arises from the fact that the timebase over which observations have
been obtained for many of the most slowly rotating stars is still
shorter than their periods. Accordingly, only lower limits of those
periods are available to date, and the incomplete phase coverage of the
magnetic data prevents the strengths and the geometrical structures of
the fields from being fully characterised.

The most famous example of this situation is $\gamma$~Equ
(=~HD~201601). The first determinations of its mean longitudinal
magnetic field $\Hz$ were achieved in 1946 by
\citet{1958ApJS....3..141B}. Since then, many additional measurements
of that field moment have been obtained by a large number of different
groups \citep[see][and references
therein]{2006MNRAS.365..585B,2015ASPC..494..100B}. They provide a
nearly continuous coverage of the $\Hz$ variations over almost 70
years, but still fall very short of completing a full cycle as the
rotation period is most likely around 100 years.

Most other very slowly rotating Ap stars have been monitored over much
shorter timespans. While prior to this study 16 were known (to us) to
have rotation periods longer than 1000 days,  only for 4 of them had
observations been obtained over more than one cycle and 
the exact value of the period been derived: HD~9996
\citep[$\Prot=7937$\,d,][]{2014AstBu..69..315M}, HD~59435
\citep[$\Prot=1360$\,d,][]{1999A&A...347..164W}, 
HD~94660 \citep[$\Prot=2800$\,d, Mathys, in preparation; see
also][]{1993ASPC...44..547H}, 
and HD~187474 \citep[$\Prot=2345$\,d,][]{1991A&AS...89..121M}. Only
lower limits were 
available for all the others, ranging from $\sim$4.5 to $\sim$100 years. 

Here we report the determination of one more period longer than 1000
days, for the star HD~18078.

HD~18078 (=~BD~+55~726) is an A0p SrCr star
\citep{2009A&A...498..961R}. On account of its low $\vsi$, it was
included by \citet{1970PASP...82..878P} in a list of 25 stars that
might have long periods, which also featured  several of the
above-mentioned slowly rotating stars (HD~9996, HD~110066, HD~187474,
and HD~201601). \citet{1971ApJ...164..309P} inferred from
consideration of differential broadening of spectral lines of
different magnetic sensitivities that HD~18078 had a magnetic field of
the order of 3.8~kG. Photometric observations by
\cite{1973PASP...85..141W} suggested that its period must
be longer than one year. Following the observation of lines resolved
into their Zeeman components in its spectrum
\citep{1992A&A...256..169M}, HD~18078 was included in the programme of
systematic study of Ap stars with magnetically resolved lines that was
led by one of our team (GM).  Its mean magnetic field modulus was found to
show large variations, but the number of measurements obtained by 1995
was not sufficient to determine the stellar rotation period
\citep{1997A&AS..123..353M}.  

We present additional determinations of that field moment in
Sect.~\ref{sec:magnetic}, and combine them with measurements of the
mean longitudinal magnetic field to derive an unambiguous value for
the rotation period. We show in Sect.~\ref{sec:photom} that this value
is consistent with observations of the photometric
variations. Finally, in Sect.~\ref{sec:discussion}, we present a
simple model of the unusual structure of the magnetic field of
HD~18078. 

\section{Magnetic variations}
\label{sec:magnetic}
\subsection{Mean magnetic field modulus}
\label{sec:modulus}
High-resolution ($R=\lambda/\Delta\lambda\simeq7\,10^4$--$1.2\,10^5$)
spectra of HD~18078 were recorded in natural light on 11 different
nights between  October 1990 and December 1997, using the AURELIE
spectrograph on the 1.52\,m telescope of Observatoire de
Haute-Provence, the Gecko spectrograph on the Canada-France-Hawaii
Telescope, and the coud\'e spectrograph with the 0.9\,m coud\'e feed
telescope of Kitt Peak National Observatory. The details of the instrumental
configurations that were used and the data reduction procedure are as
described by \citet{1997A&AS..123..353M}.  

\begin{table}
\caption{Mean longitudinal magnetic field measurements.}
\label{tab:long}
\centering
\begin{tabular}{ccrr}
\hline\hline\\[-4pt]
HJD&S/N&\multicolumn{1}{c}{$\Hz$ (G)}&$\sigma_z$ (G)\\[4pt]
\hline\\[-4pt]
2451239.173&150&$-$550&40\\
2451417.513&120&500&140\\
2451482.426&150&800&70\\
2451803.464&150&1210&70\\
2451805.541&150&910&60\\
2451806.507&150&930&70\\
2451807.484&120&900&60\\
2451862.336&120&970&90\\
2452129.515&170&40&70\\
2452153.397&120&$-$70&100\\
2452189.458&120&$-$210&70\\
2452625.321&180&$-$380&50\\
2452626.283&180&$-$410&50\\
2452689.183&180&$-$40&50\\
2452690.225&180&$-$80&40\\
2452834.529&200&810&60\\
2452917.408&160&860&70\\
2453040.410&150&990&50\\
2453273.390&200&980&60\\
2453364.179&230&620&30\\
2453666.329&180&$-$850&50\\
2453667.315&200&$-$850&40\\
2453953.545&210&$-$600&20\\
2454015.304&210&$-$200&50\\
2454162.287&180&650&50\\[4pt]
\hline
\end{tabular}
\end{table}

The \Feline\ line is resolved in its two magnetically split components
in 9 of the 11 spectra. We measured the wavelength separation of the
components to determine the mean magnetic field modulus $\Hm$ at
the corresponding epochs, by application of the formula
\begin{equation}
\lambda_{\rm r}-\lambda_{\rm b}=g\,\Zeeman\,\Hm\,.
\label{eq:Hm}
\end{equation}
In this equation, $\lambda_{\rm r}$ and $\lambda_{\rm b}$ are,
respectively, the wavelengths of the red and blue split line
components; $g$ is the Land\'e factor of the split level of the
transition ($g=2.70$; \citealt{1985aeli.book.....S}); 
$\Zeeman=k\,\lambda_0^2$, with
$k=4.67\,10^{-13}$\,\AA$^{-1}$\,G$^{-1}$; $\lambda_0=6149.258$\,\AA\
is the nominal wavelength of the considered transition.

\begin{figure}
\resizebox{\hsize}{!}{\includegraphics{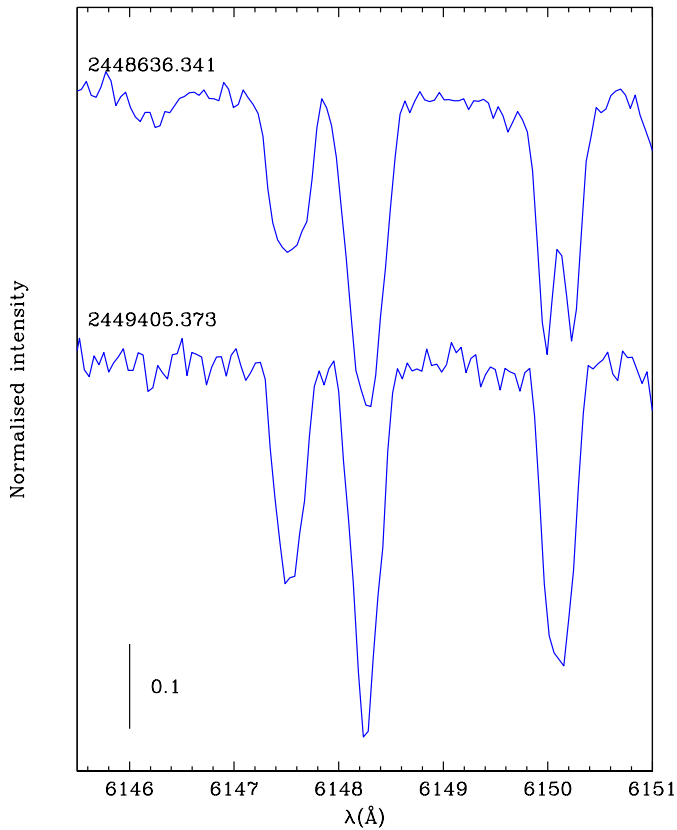}}
\caption{Portion of the spectrum of HD~18078 observed with AURELIE in
January 1992 ({\it top\/}; close to the field modulus maximum; $\Hm=4.4$\,kG) and
in February 1994 ({\it bottom\/}; when the \Feline\ line was not resolved;
$\Hm\simeq2.6$\,kG). 
The lines shown are {Cr~{\sc ii}~$\lambda\,6147.1$},
{Fe~{\sc ii}~$\lambda\,6147.7,$} and {Fe~{\sc
    ii}~$\lambda\,6149.2$}.}
\label{fig:spectrum}
\end{figure}

 In October 1993 and in February 1994,  the \Feline\ line was clearly
broadened, but not resolved.\footnote{An unrecognised radiation event
  in the middle of the line was originally mistaken for magnetic
  resolution, leading to publication of a spurious value of $\Hm$ by
  \citet{1997A&AS..123..353M} for the October 1993 observation (on
  HJD~2449286.660). That 
  measurement is superseded by the upper limit estimate presented
  here.} We estimate from its appearance that the 
mean magnetic field modulus must have been  2.6\,kG or
somewhat lower on those two dates. The \Feline\ line has been
easily resolved into its magnetically split components in similar
observations of stars with a field modulus down to 2.2\,kG
\citep{1997A&AS..123..353M}. That it cannot be resolved in HD~18078
reflects the unusually large width of its split components in that
star, which implies that the spread of the magnetic field strengths at
different locations across the stellar surface is significantly
broader than in most other stars with magnetically resolved
lines. Figure~\ref{fig:spectrum} shows a portion of the spectrum at 
two epochs, one close to the $\Hm$ maximum, the other close to its
minimum.

The derived values of $\Hm$ are listed in Table~\ref{tab:modulus},
which includes four values already published by
\citet{1997A&AS..123..353M} for the sake of completeness. As
mentioned by those authors, the measurement uncertainties are
difficult to assess. Here we revise them to 120~G, taking into account
both the difficulty of the $\Hm$ determination because of
the unusual width of the resolved line components and the scatter of the $\Hm$
data point about a sinusoidal fit (see
Sect.~\ref{sec:period}). However, this is only a rough estimate. The
two above-mentioned observations obtained at epochs when the \Feline\
line was not resolved into its magnetically split components are
flagged as ``unresolved'' in the last column of Table~\ref{tab:modulus}.

\citet{1997A&AS..123..353M} have shown that $\Hm$ measurements
obtained with different telescope and instrument combinations are, in
general, quite consistent with each other. As an exception, for a few stars
only, mean magnetic field modulus values determined from AURELIE
spectra show systematic discrepancies with respect to those obtained
with other configurations. The reason is unknown, so that we cannot
know if HD~18078 is affected by that systematic error. However,  its
magnitude does not significantly exceed 160~G for any known case: not
only it is of the same order as the estimated $\Hm$ measurement
uncertainties for HD~18078, but it is also  quite small with respect to
the amplitude of the field modulus variations in that
star. Accordingly, we do not expect it to have a significant impact on
the analysis performed in this paper. We note in particular that the
rotation period of the star, as determined in Sect.~\ref{sec:period}, is
primarily constrained by the mean longitudinal magnetic field
data, and that all field modulus values, including those derived from
AURELIE observations, define a smooth variation curve when plotted
against the phases that are computed for that value of the period.

\subsection{Mean longitudinal magnetic field}
\label{sec:long}
The Main Stellar Spectrograph of the 6\,m telescope BTA of the Special
Astrophysical Observatory was used to record medium-resolution
($R\simeq14\,500$) spectra of HD~18078 on 25 nights  from March
1999 to March 2007. The instrumental configuration and the data
reduction procedure are as described in detail by
\citet{2014AstBu..69..427R}.

The mean longitudinal magnetic field $\Hz$ was determined at each
epoch from the wavelength shifts of a sample of spectral lines between
the two circular polarisations, by application of the formula
\begin{equation}
\lambda_{\rm R}-\lambda_{\rm L}=2\,\bar g\,\Zeeman\,\Hz\,,
\label{eq:Hz}
\end{equation}
where $\lambda_{\rm R}$ (resp. $\lambda_{\rm L}$) is the wavelength of
the centre of gravity of the line in right (resp. left) circular
polarisation and $\bar g$ is the effective Land\'e factor of the
transition. The value of $\Hz$ is determined through a 
least-squares fit of the measured values of $\lambda_{\rm
  R}-\lambda_{\rm L}$ by a function of the form given above. The standard error
$\sigma_z$ that is derived from this
least-squares analysis is used as an estimate of the uncertainty
affecting the obtained value of $\Hz$. 

\begin{figure}
\resizebox{\hsize}{!}{\includegraphics{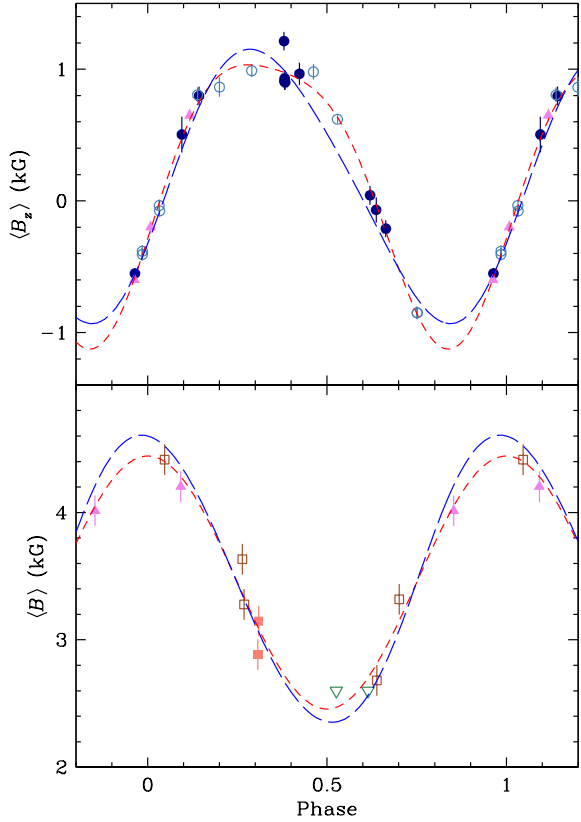}}
\caption{Mean longitudinal magnetic field ({\it top\/}) and mean
  magnetic field modulus ({\it bottom\/}) of HD~18078 against rotation
  phase. In the top panel the different symbols distinguish observations
  obtained in three different consecutive cycles (in order,
    filled dots, open circles, and filled triangles). In the bottom
  panel, open squares correspond to observations obtained with
  AURELIE, filled squares to KPNO coud\'e spectra, and filled triangles to GECKO data. The two open triangles pointing
downwards identify upper limit estimates from AURELIE spectra in which
the \Feline\ line is not magnetically resolved ({see text}). The
short-dashed lines are the best fits of the observations by a cosine wave
(for $\Hm$) and by a cosine wave and its first harmonic (for
$\Hz$). The long-dashed lines show the variations of the two
considered field moments that are predicted by the simple model
discussed in Sect.~\ref{sec:discussion}.}
\label{fig:magnetic}
\end{figure}

The values of the mean longitudinal field obtained in this way are
presented in Table~\ref{tab:long}. The signal-to-noise ratio (S/N) of
  the measured spectra  is given in Col.~2.

\subsection{Rotation period}
\label{sec:period}
Following \citet{1997A&A...320..497M}, the rotation period $\Prot$ of
HD~18078 was derived by looking for the best fit of its magnetic
variations by either a cosine wave, or by the superposition of a
cosine wave and of its first harmonic. This analysis unambiguously
indicated that $\Prot=1358$\,d. In particular, shorter periods are
definitely ruled out. The uncertainty of the derived period was
estimated by varying it between 1338\,d and 1378\,d, and by plotting for each
trial a phase diagram of the longitudinal field, in which different
symbols were used to represent the measurements corresponding to
different cycles (three of them in total). In this way, we were able
to determine 
when a significant systematic phase shift started to appear
from one cycle to the next, indicative of a value of the period that
is either too large or too small. The threshold was found at
$\pm12$\,d, so that, finally,
\begin{equation}
\Prot=(1358\pm12)\,{\rm d}.
\end{equation}

The corresponding best fits are
\begin{eqnarray}
\Hz(\phi)&=&(158\pm14)\nonumber\\
&+&(1070\pm20)\,\cos\{2\pi\,[\phi-(0.330\pm0.003)]\}\nonumber\\
&+&(217\pm20)\,\cos\{2\pi\,[2\phi-(0.196\pm0.013)]\}\,,\\
\Hm(\phi)&=&(3450\pm60)\nonumber\\
&+&(993\pm105)\,\cos\{2\pi\,[\phi-(0.000\pm0.012)]\}\,,
\end{eqnarray}
where the field strengths are expressed in Gauss, $\phi=({\rm
  HJD}-{\rm HJD}_0)/\Prot$ (mod 1), and the adopted 
value of ${\rm HJD}_0=2449930.0$ corresponds to a maximum of the mean
magnetic field modulus. The field measurements and the fitted curves
are shown in Fig.~\ref{fig:magnetic}. For the two high-resolution
spectra in which the \Feline\ line is not observationally resolved into its
magnetically split components, estimates of the upper limit of $\Hm$
at the corresponding phases are also shown. 

\begin{figure*}
\centering
\includegraphics[height=17cm,angle=270]{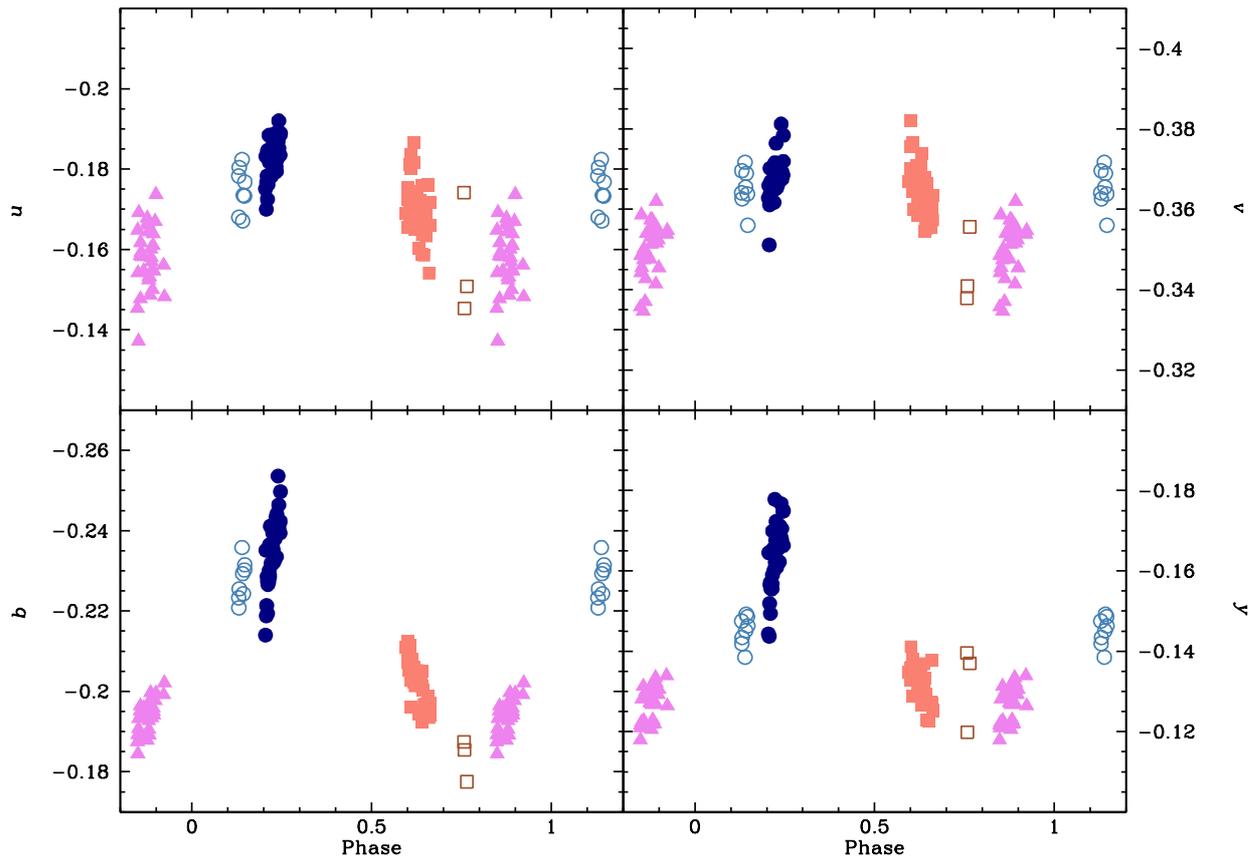}
\caption{Phase diagrams of the $uvby$ photometric measurements of
  HD~18078 for the rotation period $\Prot=1358$\,d derived from
  analysis of the magnetic variations and phase origin ${\rm
    HJD}_0=2449930.0$. Different symbols are used to 
  distinguish the observing seasons: open circles: 1995--96;
  filled squares: 1997--98; filled triangles: 1998--99;
  open squares: 2001--02; full dots: 2003--04.}
\label{fig:uvby}
\end{figure*}

 The $\Hm$ fit has been restricted to a single cosine, since the first
harmonic would not be formally significant. In contrast, the $\Hz$
variation curve very significantly departs from a single sinusoid. This is quite
unusual. While some anharmonicity of the $\Hm$ variations is
observed rather frequently, for more than 90\%\ of the Ap stars with resolved
lines for which the $\Hz$ variations have been characterised, these
variations do not show any significant departure from harmonicity. In
the few cases when such departures occur, the field modulus variations
are also anharmonic. Thus HD~18078 is  unique in showing
anharmonic variations of the longitudinal field but not of the field
modulus. 

Furthermore, HD~18078 also shows a large shift (0.17) between the
phase of the negative extremum of $\Hz$ and the phase of maximum
$\Hm$. Again, this is exceptional; in most Ap stars with magnetically
resolved lines, both curves are either almost exactly in phase or
almost exactly in antiphase. 

We return to these anomalies in Sect.~\ref{sec:discussion}.

\section{Photometric variations}
\label{sec:photom}
Str\"omgren four-colour measurements of HD~18078 were obtained using
the Four College Automatic Photometric Telescope (FCAPT) from 1995
through 2004 as part of a programme to find and improve the periods of
 magnetic chemically peculiar (mCP) stars.  The comparison star was
HD~18460 and the  check star was HD~17429.  The comparison and check stars
are near the variable on the sky and have similar $V$ magnitudes and $B-V$
colors.  They were chosen from presumably non-variable stars according
to Hipparcos photometry \citep{1997ESASP1200.....E}. The FCAPT first
measures the dark 
count and then in each filter sky-ch-c-v-c-v-c-v-c-ch-sky for each
group of variable (v), check (ch), and comparison (c) stars, where sky
is a reading of the sky.  

The FCAPT operated from 1990 through
2013. It is an automated telescope without an on-site observer,
thus the users have to be especially careful about which data to keep.
Data from groups that were not completely observed were not
analysed. In a group, if any of the standard deviations of the
differences between the comparison and check stars exceeded 2\% of the
average values, we 
excluded all the observations of the affected group  following
\citet{1988ApJS...67..439S}.  Light curve inspections still showed some
obvious outliers.  Then we compared the difference between the value
of an apparent outlier and the value of the fit at a given phase to
the standard deviation of the fit.  Any points more than 3 standard
deviations from the fit in any filter were removed from the data sets
of all filters.There were only three such outliers, two of which were only
discrepant in the $u$ and $v$ filters, that were not used in the
periodograms. For the third, the values in $b$ and $y$ were just
above the threshold for discarding data. Including these points in the
phase diagrams clearly shows that they would not change the results
from the periodograms.

The photometric data are in Table~3, available at the
CDS, which contains the following information. Column~1 lists the
heliocentric Julian Date, Cols.~2 through 9 give the variable$-$comparison magnitude (${\rm v}-{\rm c}$) and comparison$-$check
(${\rm c}-{\rm ch}$) magnitude
for the $uvby$ filters, respectively. 

The $uvby$ phase coverage of HD~18078 is limited but the variations are
consistent with the period of 1358 days determined from the magnetic
measurements (Fig.~\ref{fig:uvby}).  We used the Scargle periodogram
\citep{1982ApJ...263..835S,1986ApJ...302..757H} to search for periodic
variations in the $uvby$ 
data; one of the peaks was near 1400 days.  Although the 1995--96 and
2003--04 FCAPT data are better aligned if a period of 1405 days is
used, the $\Hz$ data, which were obtained over three cycles of
variation, show systematic shifts between consecutive cycles  when
plotted with the 
longer period and are better matched with the 1358-day period.  The
other peaks we found were for periods of 292 days and 1639 days,
neither of which is fit by the magnetic data. 

Although the phases of the photometric extrema are uncertain because
of incomplete coverage, they definitely appear shifted with
respect to those of the mean magnetic field modulus. Most likely,
brightness minimum (in all four Str\"omgren bands) occurs close to the
negative $\Hz$ extremum, and brightness maximum close to the 
positive $\Hz$ extremum, which is shallower. This suggests that the brightness
distribution over the stellar surface may to first order be symmetric
about the magnetic axis, and show a monotonic gradient from the darker 
negative magnetic pole to the brighter positive pole.

\section{Discussion}
\label{sec:discussion}
Besides the significance of the addition of a fifth star to the group
of Ap stars with exactly known rotation periods longer than 1000 days,
HD~18078 proves to be a particularly interesting object because of
the unusual structure of its magnetic field. This shows through
in a number of different ways: the magnetically split components of
the \Feline\ line are exceptionally broad; the variation curve of the
mean longitudinal magnetic field is strongly anharmonic, while that of
the mean magnetic field modulus does not show any significant
departure from a sinusoid; and the phase difference between these two
curves is closer to quadrature than to phase coincidence or 
anti-coincidence. In all these respects, the behaviour of HD~18078
is significantly different from that of the vast majority of the Ap stars
with magnetically resolved lines. 

In particular, the considerable phase shift between the $\Hm$ and
$\Hz$ variation curves implies that the magnetic field cannot be
symmetric about an axis passing through the centre of the star. Thus,
it cannot be approximated by a simple symmetric model such as
successfully used by \citet{2000A&A...359..213L} for a statistical
sample of Ap stars with magnetically resolved lines. However, the
curves of variation of $\Hm$ and $\Hz$ look separately 
rather like those of stars which can be described by the simple
centred, colinear multipole model used in that work. We consider
whether we can find a simple modification of that class of models that
reproduces at least the qualitative features of the data on HD~18078. 

We start from the view that the approximately sinusoidal variation of
$\Hz$, and the modest ratio $\Hm/\Hz \sim 3$, both suggest that there
is an important component of the field that is roughly dipolar. The
failure of the global dipolar topology for this star appears to be due
to a stronger field in the hemisphere that is coming into view as the
negative pole rotates out of sight behind the star and the positive pole
rotates into sight, compared to the weaker field that comes into view
as the positive pole of the dipole recedes behind the star. Thus, we
could consider a perturbed colinear multipole model in which the field
in one hemisphere is made weaker than  in the other hemisphere.

A simple version of this idea can be implemented by considering a
coordinate system anchored to the star, with its $z_{\rm m}$ axis parallel
to the axis of the unperturbed dipole, the $x_{\rm m}$ axis in the plane of
the stellar rotation axis and the unperturbed dipole, and the $y_{\rm m}$
axis perpendicular to the plane of the rotation axis and the
unperturbed dipole axis. It is clear that we need a perturbation in
the $y_{\rm m}$ direction, with regions of $y_{\rm m}>0$ perturbed in the
opposite sense to regions with $y_{\rm m}<0$. Within the programmes that
we use for computations of the field predicted by the centred colinear
multipole model, a perturbation of this form is very easy to
implement. One of the simplest   is to modify the
equations of the dipole, quadrupole, etc., in such a way that the field
on the surface of the star with a particular $y_{\rm m}$ coordinate is
increased or decreased by a factor $1+A\,y_{\rm m}$. Since we take the
magnetic coordinates to be normalised to the star radius, we consider
$|A|<1$.

A few experiments showed that a value of $A\approx0.4$
together with suitable polar strengths for the unperturbed dipole,
quadrupole, and octupole (which the modelling programme tries to
optimise for a good fit to the observations) is capable of reproducing
qualitatively the most extraordinary feature of the $\Hm$ and $\Hz$
curves, namely the large phase shift between the extrema of the two
curves, and--at least approximately--also fits the shapes of both
curves (see Fig.~\ref{fig:magnetic}). This supports the qualitative
discussion above and validates 
the idea that the main perturbation on the simple multipolar model
that is needed to describe the observations is to make the field in
one hemisphere of the star (with $y_{\rm m} >0$) stronger and the
field in the other hemisphere ($y_{\rm m}<0$) weaker. 

Clearly it will be possible to map this behaviour in more detail
\citep[e.g.][]{2015MNRAS.453.2163S} when
suitable observational material (polarised spectra in all four Stokes
parameters, distributed over the full rotation cycle) becomes
available. Of course, the very small value of $\vsi$ means that
HD~18078 is not remarkably suitable for mapping as there is no
velocity resolution across the disk, so such a map will probably lack
detail. Nevertheless, we expect that it will qualitatively reflect the
hemispheric asymmetry that we have identified. 

As time passes and data covering a full rotation cycle become
available for an increasing number of Ap stars with periods of  several years, it will be interesting to find out if the kind
of asymmetric magnetic field structure found in HD~18078 is more
likely to occur in extremely slowly rotating stars than in shorter period
ones, or if the conjunction of such a structure and very slow rotation
in HD~18078 is purely coincidental. 

\begin{acknowledgements}
IIR and DOK are grateful to the Russian Scientific Foundation
for partial financial support of the observations with the 6\,m telescope
and data reduction (RSF grant 14-50-00043). 
\end{acknowledgements}

\bibliographystyle{aa}
\bibliography{hd18078_rev2}
\end{document}